\documentclass[
  aps,
  pre,
  preprint,
  superscriptaddress,
  longbibliography
]{revtex4-2}

\usepackage[T1]{fontenc}
\usepackage{amsmath,amssymb,bm,mathtools}
\usepackage{booktabs}
\usepackage{graphicx}
\usepackage{pgfplots}
\usepackage{url}
\usepackage[colorlinks=true,linkcolor=blue,citecolor=blue,urlcolor=blue]{hyperref}

\pgfplotsset{compat=1.18}
\usepgfplotslibrary{groupplots}
\allowdisplaybreaks

\newcommand{\dd}{\mathrm{d}}

\begin{document}

\title{Nonreversible Gauge Fields in Fokker--Planck Dynamics: Supersymmetric Hamiltonians and Learned Finite Forces}

\author{Masayuki Ohzeki}
\email{mohzeki@tohoku.ac.jp}
\affiliation{Graduate School of Information Sciences, Tohoku University, Sendai 980-8579, Japan}
\affiliation{Department of Physics, School of Science, Institute of Science Tokyo, Tokyo 152-8551, Japan}
\affiliation{Research and Education Institute for Semiconductors and Informatics, Kumamoto University, Kumamoto 860-8555, Japan}
\affiliation{Sigma-i Co., Ltd., Tokyo 108-0075, Japan}
\date{\today}

\begin{abstract}
We formulate stationary-density-preserving nonreversible perturbations of Fokker--Planck dynamics as gauge fields that deform relaxation spectra while leaving the invariant state fixed.
When detailed balance holds, a similarity transformation maps the reversible Fokker--Planck operator to a Witten-Laplacian-type supersymmetric Hamiltonian; nonreversible gauges then appear as non-Hermitian perturbations that preserve the zero mode but modify the excited spectrum.
This operator viewpoint gives a common language for relaxation gaps, circulating probability currents, hypocoercive acceleration, and finite control costs.
We represent admissible gauge currents by antisymmetric tensor fields and identify the detailed-balance-violating Ohzeki--Ichiki force as a constant symplectic example whose infinite-strength limit is Hamiltonian dynamics.
The continuous-time spectral gap alone does not select a finite gauge strength, so we introduce a finite-time regularized objective and an actor--critic procedure for learning the gauge.
An exactly solvable anisotropic Gaussian Ornstein--Uhlenbeck benchmark separates the spectral transition from the finite-time optimum and shows that the learned gauge recovers the Lyapunov-equation optimum.
A double-well benchmark then illustrates the same constrained selection in a nonconvex metastable landscape.
Stochastic gradient methods enter this framework as physically relevant Fokker--Planck systems: mini-batch noise acts as an effective diffusion tensor, and adaptive methods such as Adam correspond to metric choices with possible nonequilibrium currents.
\end{abstract}

\maketitle

\section{Introduction}

Relaxation in stochastic dynamics is controlled not only by the stationary distribution but also by the probability currents that carry a density toward that distribution.
Detailed balance is a sufficient condition for stationarity, but it is not necessary.
Nonreversible currents that preserve the invariant density while violating detailed balance are known to accelerate relaxation in Langevin and Markov-chain dynamics, and have been studied in nonequilibrium statistical physics, stochastic sampling, and variance reduction \cite{IchikiOhzeki2013,OhzekiIchiki2015PRE,OhzekiIchiki2015JPCS,ReyBelletSpiliopoulos2016,DuncanLelievrePavliotis2016}.
The generalized Ohzeki--Ichiki dynamics provides a particularly transparent construction: a symplectic nonconservative force preserves an arbitrary Gibbs distribution, interpolates between reversible Langevin dynamics and Hamiltonian dynamics, and reaches the Hamiltonian Monte Carlo (HMC) limit at infinite gauge strength \cite{IchikiOhzeki2021}.
This interpolation, however, does not by itself determine the finite strength of the nonreversible force.
A physical principle is needed once relaxation speed, force amplitude, entropy production, discretization stability, and possible Metropolis rejection are all considered.

This paper develops an operator-theoretic framework for this finite-gauge problem.
We write Fokker--Planck dynamics as a Hamiltonian time evolution and, when detailed balance holds, map the reversible operator by a similarity transformation to a Witten-Laplacian-type supersymmetric Hamiltonian \cite{Risken1984,Witten1982}.
In this representation, relaxation is governed by low-lying energy gaps.
A stationary-density-preserving nonreversible perturbation acts as a gauge field: it leaves the zero mode, and hence the invariant density, unchanged while deforming the excited spectrum.
The central physical message is that nonreversible gauges provide a controlled way to design the relaxation spectrum of a Fokker--Planck operator without changing its stationary state.

Stochastic gradient methods provide a useful and timely class of Fokker--Planck systems to which this structure applies.
Stochastic gradient Langevin dynamics (SGLD) connects stochastic optimization and scalable Bayesian sampling \cite{WellingTeh2011}, and its finite-step behavior has been analyzed through Fokker--Planck equations and It\^o processes \cite{SatoNakagawa2014}.
Constant-step SGD can be interpreted as approximate Bayesian inference \cite{Mandt2017}, while the ratio between learning rate and batch size controls an effective noise scale \cite{Jastrzebski2017,SmithLe2018}.
In the present framework, mini-batch noise becomes an effective diffusion tensor or temperature, adaptive methods such as Adam become metric choices, and nonequilibrium deviations from reversible Langevin dynamics are organized as gauge currents.

The contributions of this paper are as follows.
First, we formulate a supersymmetric-Hamiltonian representation of reversible stochastic-gradient Fokker--Planck dynamics.
This gives a spectral language in which mini-batch noise, injected Langevin temperature, learning-rate decay, and batch-size growth are interpreted as changes of the low-lying Fokker--Planck spectrum and the stationary density.

Second, we represent stationary-density-preserving nonreversible currents by antisymmetric tensor gauge fields.
This identifies the Ohzeki--Ichiki force as a concrete constant-gauge construction and clarifies its relation to the HMC limit.
The gauge changes the excited spectrum while keeping the ground state fixed, so the problem becomes one of constrained spectral design.

Third, we show that the continuous-time spectral gap alone does not select a finite gauge strength.
We therefore introduce a finite-time regularized objective that balances relaxation against gauge-force cost, and we formulate an actor--critic procedure for learning the gauge.
An exactly solvable anisotropic Gaussian Ornstein--Uhlenbeck benchmark shows that the learned gauge matches the finite-time optimum computed from Lyapunov equations.
A nonconvex double-well benchmark then illustrates the same constrained selection in a metastable landscape.

\section{From Stochastic Gradients to Fokker--Planck Hamiltonians}
\subsection{Stochastic Gradient Methods}

Let $\mathcal{D}=\{z_n\}_{n=1}^{N}$ be a training dataset, where \(z_n\in\mathcal{Z}\) denotes the \(n\)-th data point and \(N\) is the number of samples.
Let \(\theta\in\mathbb{R}^d\) be the parameter vector to be optimized, and  $\ell:\mathbb{R}^d\times\mathcal{Z}\to\mathbb{R} $
be the loss associated with a single data point.
The empirical loss is then defined by
\begin{equation}
  L(\theta)
  =
  \frac{1}{N}
  \sum_{n=1}^{N}
  \ell(\theta;z_n).
  \label{eq:empirical-loss}
\end{equation}
Throughout the paper, \(\nabla\) denotes the gradient with respect to \(\theta\).

For a mini-batch \(B\subset\{1,\ldots,N\}\) of size \(|B|=b\), the mini-batch gradient is
\begin{equation}
  \hat g_B(\theta)
  =
  \frac{1}{b}
  \sum_{n\in B}
  \nabla \ell(\theta;z_n).
\end{equation}
We decompose it as
\begin{equation}
  \hat g_B(\theta)
  =
  \nabla L(\theta)+\xi_B(\theta),
  \qquad
  \mathbb{E}_B[\xi_B(\theta)]=0,
\end{equation}
where \(\xi_B(\theta)\) is the mini-batch gradient noise and \(\mathbb{E}_B\) denotes expectation over the random choice of the mini-batch.
Its covariance is denoted by
\begin{equation}
  \Sigma_B(\theta)
  =
  \mathbb{E}_B
  \left[
    \xi_B(\theta)\xi_B(\theta)^{\mathsf T}
  \right].
\end{equation}
With learning rate $\eta$ and preconditioner $G(\theta)$, a local approximation of SGD or Adam-type updates is
\begin{equation}
  \theta_{k+1}=\theta_k-\eta G(\theta_k)\hat g_B(\theta_k).
\end{equation}
We here use the diffusion approximation.
The diffusion approximation is obtained by matching the first two conditional moments of one discrete SGD step with those of an It\^o stochastic differential equation.
Starting from
\begin{equation}
  \theta_{k+1}
  =
  \theta_k
  -
  \eta G(\theta_k)
  \left\{
    \nabla L(\theta_k)+\xi_B(\theta_k)
  \right\},
\end{equation}
The one-step increment is
\begin{equation}
  \Delta\theta_k
  :=
  \theta_{k+1}-\theta_k
  =
  -\eta G(\theta_k)\nabla L(\theta_k)
  -
  \eta G(\theta_k)\xi_B(\theta_k).
\end{equation}
Conditioned on the current parameter value \(\theta_k=\theta\), its mean and covariance are
\begin{align}
  \mathbb{E}_B[\Delta\theta_k\mid\theta_k=\theta]
  &=
  -\eta G(\theta)\nabla L(\theta),\\
  \mathrm{Cov}_B[\Delta\theta_k\mid\theta_k=\theta]
  &=
  \eta^2
  G(\theta)\Sigma_B(\theta)G(\theta)^{\mathsf T}.
\end{align}
We now introduce the continuous time variable
\begin{equation}
  t=k\eta,
  \qquad
  \Delta t=\eta,
\end{equation}
and seek an It\^o diffusion whose increment over a short time interval \(\Delta t\) has the same leading-order mean and covariance.
For an SDE
\begin{equation}
  \dd\theta_t
  =
  b(\theta_t)\,\dd t
  +
  \sigma(\theta_t)\,\dd W_t,
\end{equation}
where \(W_t\) is a standard \(d\)-dimensional Brownian motion, the increment over \(\Delta t\) satisfies
\begin{align}
  \mathbb{E}[\Delta\theta_t\mid\theta_t=\theta]
  &=
  b(\theta)\Delta t
  +o(\Delta t),\\
  \mathrm{Cov}[\Delta\theta_t\mid\theta_t=\theta]
  &=
  \sigma(\theta)\sigma(\theta)^{\mathsf T}\Delta t
  +o(\Delta t).
\end{align}
Matching these expressions with the discrete SGD moments and using \(\Delta t=\eta\), we obtain
\begin{equation}
  b(\theta)
  =
  -G(\theta)\nabla L(\theta),
  \qquad
  \sigma(\theta)\sigma(\theta)^{\mathsf T}
  =
  \eta\,G(\theta)\Sigma_B(\theta)G(\theta)^{\mathsf T}.
\end{equation}
Thus the diffusion approximation of mini-batch SGD is
\begin{equation}
  \dd\theta_t
  =
  -G(\theta_t)\nabla L(\theta_t)\,\dd t
  +
  \sqrt{
    \eta\,G(\theta_t)\Sigma_B(\theta_t)G(\theta_t)^{\mathsf T}
  }\,
  \dd W_t^{\mathrm{mb}}.
  \label{eq:sgd-sde}
\end{equation}
Here \(W_t^{\mathrm{mb}}\) is a standard Brownian motion representing the accumulated mini-batch fluctuation, and \(\sqrt{A}\) denotes any matrix square root satisfying
\(\sqrt{A}\sqrt{A}^{\mathsf T}=A\).
This approximation is understood as a small-learning-rate, many-step limit: the deterministic part of the SGD update becomes the drift term, while the accumulated mini-batch fluctuations are treated as Brownian noise.

In stochastic gradient Langevin dynamics, an additional artificial thermal noise is added.
Using the same empirical loss \(L(\theta)\) as the potential, and denoting the target temperature by \(T>0\), we write
\begin{equation}
  \dd\theta_t
  =
  -G(\theta_t)\nabla L(\theta_t)\,\dd t
  +
  \sqrt{
    \eta\,G(\theta_t)\Sigma_B(\theta_t)G(\theta_t)^{\mathsf T}
  }\,
  \dd W_t^{\mathrm{mb}}
  +
  \sqrt{2T\,G(\theta_t)}\,\dd W_t^{\mathrm{th}},
  \label{eq:sgld-with-two-noises}
\end{equation}
where \(W_t^{\mathrm{mb}}\) and \(W_t^{\mathrm{th}}\) are independent Brownian motions.
The first noise term comes from mini-batch sampling, while the second one is the deliberately injected Langevin noise.
Equivalently, the two independent noises can be combined into a single diffusion tensor
\begin{equation}
  D_{\mathrm{eff}}(\theta)
  =
  T\,G(\theta)
  +
  \frac{\eta}{2}
  G(\theta)\Sigma_B(\theta)G(\theta)^{\mathsf T}.
  \label{eq:effective-diffusion}
\end{equation}
Here and below, \(\partial_t\) denotes differentiation with respect to time \(t\), and
\[
  \partial_i
  :=
  \frac{\partial}{\partial \theta_i}
\]
denotes differentiation with respect to the \(i\)-th component of \(\theta\).
We use the Einstein summation convention for repeated coordinate indices.

Thus the density \(\rho(\theta,t)\) of \(\theta_t\) obeys the Fokker--Planck equation
\begin{equation}
  \partial_t\rho(\theta,t)
  =
  \partial_i
  \left[
    \{G(\theta)\nabla L(\theta)\}_i
    \rho(\theta,t)
  \right]
  +
  \partial_i\partial_j
  \left[
    \{D_{\mathrm{eff}}(\theta)\}_{ij}
    \rho(\theta,t)
  \right].
  \label{eq:sgld-fp}
\end{equation}
Equivalently, this can be written as a continuity equation
\begin{equation}
  \partial_t\rho
  =
  -\nabla\cdot J,
  \qquad
  J_i
  =
  -\{G\nabla L\}_i\rho
  -
  \partial_j
  \left[
    (D_{\mathrm{eff}})_{ij}\rho
  \right].
  \label{eq:sgld-current}
\end{equation}
The vector field \(J(\theta,t)\) is the probability current associated with the density \(\rho(\theta,t)\).
When \(T=0\), this equation reduces to the diffusion approximation of mini-batch SGD.
When the mini-batch covariance term is neglected, it reduces to the usual Langevin Fokker--Planck equation with temperature \(T\).

\subsection{Supersymmetric Hamiltonian}

We now rewrite the Fokker--Planck equation obtained above as a quantum-mechanical time evolution.
Equation \eqref{eq:sgld-fp} can be written as
\begin{equation}
  \partial_t\rho
  =
  -\mathcal{H}_{\mathrm{FP}}\rho,
  \label{eq:fp-hamiltonian-time-evolution}
\end{equation}
where the Fokker--Planck Hamiltonian is the differential operator
\begin{equation}
  \mathcal{H}_{\mathrm{FP}}
  =
  -
  \partial_i
  \left[
    \{G\nabla L\}_i\,\cdot
  \right]
  -
  \partial_i\partial_j
  \left[
    (D_{\mathrm{eff}})_{ij}\,\cdot
  \right].
  \label{eq:sgld-fp-hamiltonian}
\end{equation}
The dot indicates the argument on which the operator acts.
At this stage no similarity transformation has been performed.
Therefore \(\mathcal{H}_{\mathrm{FP}}\) is generally non-Hermitian as an operator on
\(L^2(\mathbb{R}^d,\dd\theta)\), the space of square-integrable functions with respect to the Lebesgue measure \(\dd\theta\).

If an invariant density \(\pi(\theta)\) exists, it satisfies
\begin{equation}
  \mathcal{H}_{\mathrm{FP}}\pi=0.
\end{equation}
Thus \(\pi\) is the right zero eigenfunction of \(\mathcal{H}_{\mathrm{FP}}\), while probability conservation implies that the constant function \(1\) is the left zero eigenfunction.
The relaxation rate is controlled by the nonzero eigenvalues of \(\mathcal{H}_{\mathrm{FP}}\).
We define the spectral gap by
\begin{equation}
  \Delta
  =
  \min_{\lambda\in\mathrm{Spec}(\mathcal{H}_{\mathrm{FP}})\setminus\{0\}}
  \mathrm{Re}\,\lambda,
  \label{eq:sgld-fp-gap}
\end{equation}
assuming that the zero eigenvalue is isolated and that the real parts of all nonzero eigenvalues are positive.
The detailed balance with respect to \(\pi\) then means that the stationary probability current vanishes:
\begin{equation}
  J_i^{\pi}
  =
  -\{G\nabla L\}_i\pi
  -
  \partial_j
  \left[
    (D_{\mathrm{eff}})_{ij}\pi
  \right]
  =
  0.
  \label{eq:sgld-detailed-balance}
\end{equation}
By contrast, preserving the stationary density only requires the weaker stationarity condition
\begin{equation}
  \partial_iJ_i^{\pi}=0.
  \label{eq:sgld-stationarity-current}
\end{equation}
Thus a nonzero divergence-free stationary current can break detailed balance without changing the invariant density.
This stationary-density-preserving current is the gauge degree of freedom that will later be represented by antisymmetric tensors.
The Ohzeki--Ichiki force is a concrete construction of such a detailed-balance-breaking current for Langevin dynamics \cite{OhzekiIchiki2015PRE,OhzekiIchiki2015JPCS}.

We now specialize the SGLD Fokker--Planck Hamiltonian to the reversible scalar-temperature approximation.
In general, \(D_{\mathrm{eff}}(\theta)\) is state-dependent and anisotropic.
To obtain the standard supersymmetric representation without additional geometric correction terms, we assume that \(G\) is constant, symmetric, and positive definite, and approximate the mini-batch covariance by
\begin{equation}
  G\Sigma_B(\theta)G^{\mathsf T}
  \simeq
  \frac{\sigma_G^2}{b}\,G,
  \qquad
  b=|B|.
\end{equation}
Here \(b\) is the mini-batch size and \(\sigma_G^2\) is the gradient-noise strength measured in the metric defined by \(G\).
Then
\begin{equation}
  D_{\mathrm{eff}}
  \simeq
  T_{\mathrm{eff}}G,
  \qquad
  T_{\mathrm{eff}}
  =
  T+\frac{\eta\sigma_G^2}{2b}.
  \label{eq:effective-temperature}
\end{equation}
For plain mini-batch SGD without injected Langevin noise, \(T=0\); for the Euclidean case \(G=I\), this reduces to
\(T_{\mathrm{eff}}\simeq \eta\sigma^2/(2b)\).

In this limit, the stationary density satisfying detailed balance is
\begin{equation}
  \pi(\theta)
  =
  \frac{1}{Z}
  \exp\left[
    -\frac{L(\theta)}{T_{\mathrm{eff}}}
  \right],
  \qquad
  Z
  =
  \int_{\mathbb{R}^d}
  \exp\left[
    -\frac{L(\theta)}{T_{\mathrm{eff}}}
  \right]\dd\theta.
  \label{eq:sgld-gibbs-density}
\end{equation}
The similarity transformation
\begin{equation}
  \rho(\theta,t)
  =
  \pi(\theta)^{1/2}\psi(\theta,t)
\end{equation}
leads to 
\begin{equation}
  \partial_t\psi
  =
  -H_{\mathrm{SUSY}}\psi,
  \qquad
  H_{\mathrm{SUSY}}
  :=
  \pi^{-1/2}
  \mathcal{H}_{\mathrm{FP}}^{\mathrm{rev}}
  \pi^{1/2}.
  \label{eq:susy-similarity-transform}
\end{equation}
Thus, in the reversible case, the similarity-transformed Fokker--Planck Hamiltonian is the supersymmetric Hamiltonian.

For constant \(G\), its scalar, or \(0\)-form, part is
\begin{equation}
  H_{\mathrm{SUSY}}
  =
  -T_{\mathrm{eff}}
  \partial_i
  \left(
    G_{ij}\partial_j
  \right)
  +
  \frac{1}{4T_{\mathrm{eff}}}
  \partial_iL\,G_{ij}\partial_jL
  -
  \frac{1}{2}
  \partial_i
  \left[
    G_{ij}\partial_jL
  \right].
  \label{eq:susy-hamiltonian}
\end{equation}
Equivalently,
\begin{equation}
  H_{\mathrm{SUSY}}
  =
  A_i^\dagger G_{ij}A_j,
  \qquad
  A_i
  =
  \sqrt{T_{\mathrm{eff}}}
  \left(
    \partial_i
    +
    \frac{\partial_iL}{2T_{\mathrm{eff}}}
  \right),
\end{equation}
where \(A_i^\dagger\) denotes the adjoint of \(A_i\) in \(L^2(\mathbb{R}^d,\dd\theta)\).
The ground state is
\begin{equation}
  \psi_0(\theta)
  =
  \pi(\theta)^{1/2}
  \propto
  \exp\left[
    -\frac{L(\theta)}{2T_{\mathrm{eff}}}
  \right],
  \qquad
  H_{\mathrm{SUSY}}\psi_0=0.
\end{equation}
This is the Witten-Laplacian-type supersymmetric Hamiltonian associated with the loss landscape \(L(\theta)\) in the reversible SGLD approximation
\cite{Risken1984,Witten1982}.

The advantage of this transformation is that relaxation becomes an energy-gap problem.
Since \(H_{\mathrm{SUSY}}\) is Hermitian and nonnegative in the reversible case, the first nonzero eigenvalue
\begin{equation}
  \Delta
  =
  \min_{\lambda\in\mathrm{Spec}(H_{\mathrm{SUSY}})\setminus\{0\}}
  \lambda
\end{equation}
controls the leading relaxation mode:
\begin{equation}
  \rho(\theta,t)-\pi(\theta)
  \simeq
  a_1 e^{-\Delta t}\pi(\theta)^{1/2}\psi_1(\theta),
\end{equation}
where \(\psi_1\) is the first excited eigenfunction and \(a_1\) is its projection coefficient determined by the initial density.

In metastable loss landscapes, such as double-well-like landscapes, the low-temperature gap is exponentially small because transitions between basins require barrier crossing.
If \(\Delta L_{\mathrm b}\) denotes the loss barrier height between two basins, then the Eyring--Kramers form gives
\begin{equation}
  \Delta(T_{\mathrm{eff}})
  \simeq
  A_{\mathrm{EK}}
  \exp\left[
    -\frac{\Delta L_{\mathrm b}}{T_{\mathrm{eff}}}
  \right],
  \label{eq:ek-gap-scaling}
\end{equation}
up to a prefactor \(A_{\mathrm{EK}}\) determined by the local curvatures near the minimum and the saddle point \cite{BonyLePeutrecMichel2022}.
Combining \eqref{eq:effective-temperature} with \eqref{eq:ek-gap-scaling}, mini-batch noise increases the effective temperature and can enlarge the first nonzero Fokker--Planck eigenvalue in metastable regimes.
This is the precise sense in which mini-batch noise opens a relaxation gap.
The more accurate statement beyond the scalar approximation is that SGD carries the anisotropic diffusion tensor \(D_{\mathrm{eff}}(\theta)\), which can promote barrier crossing differently along different directions.
This connects with fluctuation--dissipation and effective-noise analyses of SGD, as well as studies of anisotropic gradient noise and flat-solution selection
\cite{Yaida2019,MignaccoUrbani2022,Zhu2019,DaiZhu2021,YangTangTu2023}.

It is important, however, not to confuse increasing the gap with optimizing the learning objective.
Increasing \(T_{\mathrm{eff}}\) improves mixing across barriers, but the stationary density
\(\pi(\theta)\propto\exp[-L(\theta)/T_{\mathrm{eff}}]\)
becomes less concentrated near low-loss regions.
If \(L_\star\) is the loss at a local minimum and \(d_{\mathrm{eff}}\) is the effective quadratic dimension of that basin, equipartition gives the local estimate
\begin{equation}
  \mathbb{E}_{\pi}
  \left[
    L(\theta)-L_\star
  \right]
  \simeq
  \frac{d_{\mathrm{eff}}}{2}T_{\mathrm{eff}}.
  \label{eq:equipartition-loss}
\end{equation}
Thus a useful temperature is finite.
To escape an undesirable basin with barrier height \(\Delta L_{\mathrm{esc}}\) within an exploration time \(\tau_{\mathrm{exp}}\), a Kramers estimate gives
\begin{equation}
  T_{\mathrm{eff}}
  \gtrsim
  \frac{\Delta L_{\mathrm{esc}}}
       {\log(A_{\mathrm{esc}}\tau_{\mathrm{exp}})}.
  \label{eq:temperature-lower-bound}
\end{equation}
Here \(A_{\mathrm{esc}}\) is the Kramers prefactor for escape.
On the other hand, to keep the expected excess loss below a tolerance \(\varepsilon\), \eqref{eq:equipartition-loss} gives
\begin{equation}
  T_{\mathrm{eff}}
  \lesssim
  \frac{2\varepsilon}{d_{\mathrm{eff}}}.
  \label{eq:temperature-upper-bound}
\end{equation}
Similarly, to remain in a favorable basin with barrier height \(\Delta L_{\mathrm{hold}}\) for a time scale \(\tau_{\mathrm{hold}}\), one needs approximately
\begin{equation}
  T_{\mathrm{eff}}
  \lesssim
  \frac{\Delta L_{\mathrm{hold}}}
       {\log(A_{\mathrm{hold}}\tau_{\mathrm{hold}})},
  \label{eq:temperature-retention-bound}
\end{equation}
where \(A_{\mathrm{hold}}\) is the corresponding Kramers prefactor.
Therefore, a useful effective temperature is neither obtained by maximizing the gap alone nor by increasing \(T_{\mathrm{eff}}\) without bound.
The temperature should be large enough to assist exploration and barrier crossing, but small enough to concentrate the stationary density near low-loss regions.
This interpretation has a concrete algorithmic consequence.
Since
\[
  T_{\mathrm{eff}}
  =
  T+\frac{\eta\sigma_G^2}{2b},
\]
learning-rate decay and batch-size growth both reduce the same effective diffusion scale.
Thus they are not independent heuristics in the SGLD/Fokker--Planck view; they are two ways of cooling the same diffusion process.
This is the Fokker--Planck analogue of simulated-annealing schedules, although real SGD noise is generally anisotropic and state dependent
\cite{GemanGeman1984,Hajek1988}.
This observation is consistent with the noise-scale analysis of SGD, where the relevant scale is controlled by the ratio of learning rate to batch size
\cite{Jastrzebski2017,SmithLe2018}.
It is also consistent with the empirical finding that learning-rate decay can often be replaced by increasing the batch size during training, while preserving similar training and test curves
\cite{SmithKindermansYingLe2018}.
From the present viewpoint, such a schedule gradually lowers \(T_{\mathrm{eff}}\): early training keeps enough noise to explore and escape sharp or metastable regions, whereas late training reduces the noise so that the dynamics can concentrate near low-loss regions.
This should not be read as a claim that smaller batches or larger noise are always better; rather, the useful quantity is the scheduled noise scale.
Indeed, carefully controlled experiments show that stochastic-gradient noise can improve generalization in some regimes, but that its benefit depends on the training budget and hyperparameter tuning
\cite{SmithElsenDe2020}.

The main point of the SUSY-Hamiltonian representation is not merely to rederive known noise-scale heuristics.
Rather, it provides a unified spectral language in which several empirical training practices can be compared on the same footing.
Mini-batch noise, injected Langevin temperature, learning-rate decay, and batch-size growth all change the effective diffusion tensor and hence the low-lying spectrum of the Fokker--Planck operator.
In the reversible scalar-temperature limit, this spectrum is equivalently the spectrum of the supersymmetric Hamiltonian \(H_{\mathrm{SUSY}}\).
Thus exploration, metastable escape, concentration near low-loss regions, and annealing schedules can be discussed in terms of how they modify the relaxation gap and the stationary density.
This unified viewpoint is the first benefit of the SUSY formulation: it turns a collection of optimizer heuristics into spectral statements about a single operator.

\section{Adam-Type Adaptive Learning Rates as Gauges}

\subsection{Adaptive Learning Rates}
Adam-type methods \cite{KingmaBa2015} introduce an extended state with first and second gradient moments,
\begin{equation}
  m_{t+1}=\beta_1m_t+(1-\beta_1)\hat g_t,\qquad
  v_{t+1}=\beta_2v_t+(1-\beta_2)\hat g_t^2,
\end{equation}
and update
\begin{equation}
  \theta_{t+1}=\theta_t-\eta
  \frac{\hat m_t}{\sqrt{\hat v_t}+\epsilon}.
\end{equation}
Under a continuous-time and adiabatic approximation, this is a Langevin-type equation with diagonal preconditioner
\begin{equation}
  G_{ii}(\theta)\simeq(\sqrt{v_i(\theta)}+\epsilon)^{-1}.
\end{equation}
Parameter space, therefore, no longer looks Euclidean; it carries a position-dependent metric $G$.

If one wants an adaptive-preconditioned process to be an exact reversible sampler for a fixed Gibbs density, the drift and diffusion must be matched.
For a target density \(\pi(\theta)\propto\exp[-L(\theta)/T]\), this requires the thermal noise associated with the metric \(G\) and the It\^o correction \(T\nabla\cdot G\).
With this correction, the quantum-mechanical operator replaces the ordinary Laplacian by a metric-dependent Laplace operator, and geometric connection terms enter the transformed Hamiltonian.
Standard Adam is not this reversible sampler.
An Adam-like reversible sampler is possible only in the more specific sense that a diagonal, data-adaptive preconditioner is coupled to the matching noise and correction terms, as in preconditioned or Riemannian Langevin dynamics \cite{GirolamiCalderhead2011,PattersonTeh2013,MaChenFox2015}.
If the metric is estimated online from stochastic gradients, exact reversibility is recovered only in an enlarged Markov state with a consistent fluctuation--dissipation structure, or after a Metropolis correction.
Otherwise, the adaptive metric and auxiliary moment variables generate nonequilibrium currents.
Decomposing those currents into gauge fields allows one to analyze Adam's speed as a mixture of curvature preconditioning and circulating nonreversible flow.

Thus Adam should not be viewed only as a coordinate-wise learning-rate rule.
In the continuous-time Fokker--Planck description, it is more naturally interpreted as a choice of a metric, or preconditioner, together with possible nonequilibrium currents.
In Euclidean coordinates, we write this general metric--gauge Langevin form as
\begin{equation}
  \dd\theta_t
  =
  \left[
    -G(\theta_t)\nabla L(\theta_t)
    +
    T\,\nabla\cdot G(\theta_t)
    +
    c(\theta_t)
  \right]\dd t
  +
  \sqrt{2T}\,E(\theta_t)\dd W_t,
  \qquad
  G(\theta)=E(\theta)E(\theta)^{\mathsf T}.
  \label{eq:metric-gauge-sde}
\end{equation}
Here \(G(\theta)\) is a symmetric positive semidefinite preconditioning matrix,
\(E(\theta)\) is any matrix square root of \(G(\theta)\), and
\[
  \{\nabla\cdot G(\theta)\}_i
  :=
  \partial_j G_{ij}(\theta).
\]
The vector field \(c(\theta)\) represents a nonreversible drift.
If it satisfies
\begin{equation}
  \partial_i\{\pi(\theta)c_i(\theta)\}=0,
  \label{eq:gauge-stationarity-condition}
\end{equation}
then it preserves the stationary density \(\pi(\theta)\), although it generally breaks detailed balance.

The idea of adding stationary-density-preserving currents is standard in nonequilibrium statistical physics: detailed balance is sufficient for stationarity, but it is not necessary.
A nonzero stationary current can break detailed balance while preserving the same invariant density.
In the stochastic-gradient MCMC literature, Ma, Chen and Fox recast this idea as a general recipe for constructing samplers from a positive semidefinite diffusion matrix and an antisymmetric matrix field \cite{MaChenFox2015}.
Nonreversible Langevin samplers have also been studied as a systematic way to improve convergence and reduce variance
\cite{ReyBelletSpiliopoulos2016,DuncanLelievrePavliotis2016}.
In the stochastic-gradient setting, detailed-balance violation has also been introduced into stochastic gradient Langevin dynamics to accelerate stochastic dynamics
\cite{Ohzeki2016AcceleratedSGD}.

In the notation used here, a stationary-density-preserving nonreversible drift can be generated from an antisymmetric tensor field \(A(\theta)\) as
\begin{equation}
  c_i(\theta)
  =
  \frac{1}{\pi(\theta)}
  \partial_j
  \left[
    A_{ij}(\theta)\pi(\theta)
  \right],
  \qquad
  A_{ij}(\theta)=-A_{ji}(\theta).
  \label{eq:antisymmetric-gauge-drift}
\end{equation}
This form automatically satisfies \eqref{eq:gauge-stationarity-condition}, since
\begin{equation}
  \partial_i\{\pi(\theta)c_i(\theta)\}
  =
  \partial_i\partial_j
  \left[
    A_{ij}(\theta)\pi(\theta)
  \right]
  =
  0.
\end{equation}
Thus \(A(\theta)\) is a gauge potential for nonreversible probability flow: it changes the stationary current without changing the stationary density.
The Ohzeki--Ichiki force is a concrete physical construction of such a detailed-balance-breaking current for Langevin dynamics.

\subsection{Role in Stochastic Optimization}

The metric--gauge form above also clarifies why modern stochastic optimizers exhibit different relaxation and generalization behavior.
Adam, RMSProp, AdaGrad, AdamW, Adafactor, Sophia, and Lion all use adaptive coordinate-wise or curvature-informed scaling to reduce local anisotropy in the effective landscape \cite{KingmaBa2015,LoshchilovHutter2019,ShazeerStern2018,LiuSophia2023,ChenLion2023}.
In the present notation, these methods primarily modify the metric \(G\) and, when their auxiliary variables are treated dynamically, can also generate nonequilibrium currents.
This explains why adaptive methods often accelerate early training in strongly anisotropic problems, while also making clear that a standard Adam update is not automatically a reversible Langevin sampler.

The same formulation also separates metric adaptation from the temperature-like effect of SGD noise.
Large-batch generalization gaps, sharp minima, noise-scale arguments, and empirical gradient-noise-scale laws all concern the effective diffusion tensor in \eqref{eq:sgd-sde} \cite{KeskarLargeBatch2017,Jastrzebski2017,SmithLe2018,McCandlish2018,SmithElsenDe2020}.
Switching or bounding adaptive methods, as in SWATS and AdaBound, can be viewed as changing the balance between early metric acceleration and later SGD-like diffusion \cite{KeskarSocher2017,Luo2019}.
Decoupled weight decay in AdamW prevents the adaptive metric from unintentionally deforming the radial confining part of the dynamics \cite{LoshchilovHutter2019}.
Failures or different solution selection by adaptive methods, including the convergence issues motivating AMSGrad, reflect the fact that a rapidly changing metric need not satisfy a fluctuation--dissipation relation for a fixed target density \cite{Wilson2017,Reddi2018}.

High-learning-rate phenomena such as the edge of stability are related but not identical to the continuous-time Fokker--Planck limit.
The observed stability boundary in full-batch gradient descent and the stochastic sharpness gap in mini-batch SGD require discrete-time effects beyond the small-step diffusion approximation used here \cite{Cohen2021,LiaoKolomvakiKyrillidis2026}.
Similarly, Sharpness-Aware Minimization explicitly modifies the effective landscape by penalizing local sharpness, whereas mini-batch SGD does so implicitly through diffusion \cite{Foret2021}.
These developments support the operator viewpoint without being reduced to it: metric, temperature, and discrete-time stability are distinct mechanisms.

Our focus in the remainder of the paper is the third mechanism, the stationary-density-preserving gauge freedom.
After a metric and diffusion tensor have been specified, an antisymmetric gauge current can accelerate relaxation without changing the target density.
This is the part of the dynamics that is optimized below.

\section{Gauge Fields Break Detailed Balance}

\subsection{Antisymmetric tensor field}

A stationary-density-preserving nonreversible drift can be generated by an antisymmetric tensor field.
To avoid confusion with the probability current \(J\), we denote this tensor by \(A(\theta)\).
Let
\[
  A_{ij}(\theta)=-A_{ji}(\theta).
\]
Then
\begin{equation}
  c_i(\theta)
  =
  \frac{1}{\pi(\theta)}
  \partial_j
  \left[
    A_{ij}(\theta)\pi(\theta)
  \right].
  \label{eq:antisymmetric-gauge}
\end{equation}
Equivalently, the stationary nonreversible current
\[
  J_i^\pi(\theta)=\pi(\theta)c_i(\theta)
\]
is written as
\begin{equation}
  J_i^\pi(\theta)
  =
  \partial_j
  \left[
    A_{ij}(\theta)\pi(\theta)
  \right].
\end{equation}
Its divergence vanishes identically:
\begin{equation}
  \partial_iJ_i^\pi
  =
  \partial_i\partial_j
  \left[
    A_{ij}(\theta)\pi(\theta)
  \right]
  =
  0,
\end{equation}
because \(A_{ij}\) is antisymmetric while \(\partial_i\partial_j\) is symmetric.
Thus \(A(\theta)\) is a gauge potential for stationary probability flow.

In two dimensions, the antisymmetric tensor is represented by a single scalar stream function \(\chi(\theta_1,\theta_2)\).
Writing
\begin{equation}
  A(\theta)\pi(\theta)
  =
  \begin{pmatrix}
    0 & \chi(\theta)\\
    -\chi(\theta) & 0
  \end{pmatrix},
\end{equation}
we obtain
\begin{equation}
  J^\pi(\theta)
  =
  \pi(\theta)c(\theta)
  =
  \begin{pmatrix}
    \partial_2\chi(\theta)\\
    -\partial_1\chi(\theta)
  \end{pmatrix},
  \label{eq:2d-stream}
\end{equation}
up to the sign convention used for \(A_{12}\).
This is the usual stream-function representation of a divergence-free current.

In an arbitrary dimension, the same construction is the Hodge representation of a divergence-free vector field.
If
\[
  J_i^\pi(\theta)=\pi(\theta)c_i(\theta)
\]
satisfies \(\partial_iJ_i^\pi=0\), then locally it can be represented by an antisymmetric tensor potential:
\begin{equation}
  J_i^\pi(\theta)
  =
  \partial_j\chi_{ij}(\theta),
  \qquad
  \chi_{ij}(\theta)=-\chi_{ji}(\theta).
  \label{eq:hodge-current}
\end{equation}
One may further write
\begin{equation}
  \chi_{ij}
  =
  \partial_i\Omega_j-\partial_j\Omega_i,
  \label{eq:hodge-potential}
\end{equation}
with a gauge condition such as \(\partial_i\Omega_i=0\).
Then \(\Omega\) solves a Poisson-type equation whose source is the stationary current \(J^\pi\).
This is the higher-dimensional analogue of the two-dimensional stream function.

In the Hamiltonian representation, the nonreversible gauge has a particularly transparent meaning.
The reversible Fokker--Planck operator is mapped by a similarity transformation to the Hermitian SUSY Hamiltonian \(H_{\mathrm{SUSY}}\).
A stationary-density-preserving gauge field \(A\) adds a non-Hermitian perturbation:
\[
  \widetilde{\mathcal H}_{\mathrm{FP}}
  =
  H_{\mathrm{SUSY}}
  +
  \widetilde{\mathcal H}_A .
\]
Because the gauge preserves the target density, the zero mode is unchanged.
Thus the gauge acts by deforming the excited spectrum while keeping the ground state fixed.
In this sense, gauge optimization is a spectral-design problem: one changes the low-lying relaxation modes without changing the stationary distribution.
For reversible dynamics the spectrum is real and relaxation is purely dissipative, whereas a nonreversible gauge can produce complex eigenvalues whose imaginary parts represent circulating probability currents.
In particular, Ichiki and Ohzeki showed that detailed-balance violation can generate imaginary components in the relaxation spectrum \cite{IchikiOhzeki2013}.
The Ohzeki--Ichiki symplectic force is a concrete example of such a perturbation.
They further showed that, in the strong-gauge limit, the Hamiltonian, energy-preserving part dominates and the dynamics approaches the HMC limit \cite{IchikiOhzeki2021}.
This spectral viewpoint clarifies why a finite gauge may be preferable: one wants to increase the real parts of the slow relaxation modes without paying excessive control cost or collapsing into purely energy-shell motion.

\section{Variational Principle for the Optimal Gauge}

The Ohzeki--Ichiki method shows that adding a nonreversible force while preserving the stationary distribution can accelerate relaxation.
However, it does not determine how strong the force should be.
In continuous time, an excessively strong force carries high entropy production or control cost; in discrete time, it may also worsen numerical error and Metropolis rejection.
The optimal gauge is therefore not the strongest gauge in isolation, but the gauge that gives the best balance between finite-time relaxation and a specified cost.
This specification is essential: without a force, entropy-production, discretization, or rejection constraint, there is generally no parameter-free finite strength to derive.

We formulate this balance as a finite-time transport problem in parameter space.
Let \(\rho_0(\theta)\) be the initial density of the stochastic parameters and let \(\pi(\theta)\) be the target stationary density.
For a given antisymmetric gauge field \(A(\theta)\), let \(\rho_A(\theta,t)\) be the density path generated by the Fokker--Planck dynamics with the corresponding drift \(c_A(\theta)\).
The relevant question is how efficiently \(\rho_A(\cdot,t)\) is transported from \(\rho_0\) toward \(\pi\) over a finite time horizon \([0,\tau]\).

The Benamou--Brenier formulation gives a useful reference for this transport viewpoint \cite{BenamouBrenier2000}.
If arbitrary velocity fields were allowed, the least-action density path between the prescribed endpoints would be obtained by minimizing
\begin{equation}
  \inf_{\rho,v}
  \int_0^\tau
  \int_{\mathbb{R}^d}
  \frac{1}{2}\rho(\theta,t)|v(\theta,t)|^2
  \,\dd\theta\,\dd t,
  \qquad
  \partial_t\rho+\partial_i(\rho v_i)=0,
  \label{eq:bb}
\end{equation}
with endpoint conditions
\begin{equation}
  \rho(\theta,0)=\rho_0(\theta),
  \qquad
  \rho(\theta,\tau)=\pi(\theta).
\end{equation}
This problem defines the least-action transport from the initial density to the target density.
It is also related to the Wasserstein gradient-flow interpretation of the reversible Fokker--Planck equation \cite{JordanKinderlehrerOtto1998}.

The gauge problem is more constrained.
The actual Fokker--Planck dynamics cannot realize an arbitrary velocity field.
With a gauge drift \(c_A\), the admissible current velocity is
\begin{equation}
  v_{A,i}[\rho]
  =
  v_{0,i}[\rho]+c_{A,i},
  \qquad
  v_{0,i}[\rho]
  =
  -\{G\nabla L\}_i
  -
  \frac{1}{\rho}
  \partial_j
  \left[
    (D_{\mathrm{eff}})_{ij}\rho
  \right].
  \label{eq:admissible-gauge-velocity}
\end{equation}
The density path \(\rho_A\) is therefore determined by
\begin{equation}
  \partial_t\rho_A
  +
  \partial_i
  \left[
    \rho_A v_{A,i}[\rho_A]
  \right]
  =
  0,
  \qquad
  \rho_A(\theta,0)=\rho_0(\theta).
  \label{eq:gauge-induced-density-path}
\end{equation}
One may interpret the gauge design as a proximal projection of an ideal transport flow onto this admissible family.
However, computing the Benamou--Brenier velocity is itself a hard optimal-transport problem.
We therefore use it only as motivation for the density-transport viewpoint and define the practical gauge objective directly in terms of measurable finite-time relaxation.

Let \(\mathcal R_\tau[\rho_A,\pi]\) be a nonnegative relaxation loss that measures how close the density path \(\rho_A\) comes to the target \(\pi\) during the time interval \([0,\tau]\).
The regularized gauge objective is
\begin{equation}
  \mathcal J_\lambda[A]
  =
  \mathcal R_\tau[\rho_A,\pi]
  +
  \frac{\lambda}{2}
  \int_0^\tau
  \int_{\mathbb{R}^d}
  \rho_A(\theta,t)|c_A(\theta)|^2
  \,\dd\theta\,\dd t.
  \label{eq:practical-gauge-objective}
\end{equation}
where \(\lambda>0\) is the prescribed cost coefficient for the gauge force.
For sampling, one may choose
\[
  \mathcal R_\tau[\rho_A,\pi]
  =
  D_{\mathrm{KL}}(\rho_A(\cdot,\tau)\|\pi)
\]
or its time-integrated version.
For observable relaxation, one may choose
\[
  \mathcal R_\tau[\rho_A,\pi]
  =
  \int_0^\tau
  \left|
    \mathbb E_{\rho_A(t)}[f]
    -
    \mathbb E_{\pi}[f]
  \right|
  \dd t.
\]
In the double-well experiment as shown below, the target distribution is symmetric and
\(\mathbb E_\pi[x]=0\).
Therefore we use
\begin{equation}
  \mathcal R_\tau[\rho_A,\pi]
  =
  \int_0^\tau
  \left|
    \mathbb E_{\rho_A(t)}[x]
  \right|
  \dd t.
\end{equation}
Thus the learning algorithm does not require solving the Benamou--Brenier problem.
It only requires simulating, or otherwise propagating, the Fokker--Planck dynamics under candidate gauge fields and estimating the objective in \eqref{eq:practical-gauge-objective}.

\subsection{Learning Continuous Gauge Fields by Actor--Critic Optimization}

The objective \eqref{eq:practical-gauge-objective} suggests a learning-based route to gauge design.
In general, the optimal antisymmetric tensor field \(A^\star(\theta)\) is not available in closed form.
We therefore parameterize the admissible gauge and optimize its parameters using finite-time information generated by the controlled dynamics.

We use the term actor--critic in analogy with policy-gradient methods in reinforcement learning
\cite{SuttonMcAllesterSinghMansour2000,KondaTsitsiklis2000}.
In the standard actor--critic setting, the actor is a parameterized policy, while the critic estimates the value or gradient information needed to update that policy.
In the present continuous-time setting, the actor is not a discrete action policy.
It is the parameterized antisymmetric gauge field \(A_\psi\).
The critic estimates the finite-time objective \(\mathcal J_\lambda[\psi]\) and, when needed, its gradient with respect to the actor parameters \(\psi\).
This estimate may be obtained from rollouts, as in the numerical experiment below, or from an adjoint Fokker--Planck equation.

Thus the actor is an antisymmetric tensor field
\[
  A_\psi(\theta)=-A_\psi(\theta)^{\mathsf T},
\]
with trainable parameters \(\psi\).
It generates the nonreversible drift
\begin{equation}
  c_{\psi,i}(\theta)
  =
  \frac{1}{\pi(\theta)}
  \partial_j
  \left[
    A_{\psi,ij}(\theta)\pi(\theta)
  \right].
  \label{eq:learned-continuous-gauge}
\end{equation}
This hard parameterization preserves the target density in continuous time, because
\[
  \partial_i\{\pi(\theta)c_{\psi,i}(\theta)\}
  =
  \partial_i\partial_j
  \left[
    A_{\psi,ij}(\theta)\pi(\theta)
  \right]
  =
  0.
\]
The normalization constant of \(\pi\) is not needed.
Thus learning changes the stationary current, not the target density.

Let \(\rho_\psi(\theta,t)\) be the density path generated by the Fokker--Planck dynamics with gauge drift \(c_\psi\).
Then the practical objective is \(\mathcal J_\lambda[A_\psi]\), the parameterized version of \eqref{eq:practical-gauge-objective}.
For observable relaxation, a convenient terminal choice of \(\mathcal R_\tau\) is
\begin{equation}
  \mathcal{R}_\tau[\rho_\psi,\pi]
  =
  \frac{1}{2}
  \sum_a w_a
  \left[
    \int_{\mathbb{R}^d}
    f_a(\theta)\rho_\psi(\theta,\tau)\,\dd\theta
    -
    \int_{\mathbb{R}^d}
    f_a(\theta)\pi(\theta)\,\dd\theta
  \right]^2,
  \qquad
  w_a\ge 0,
  \label{eq:finite-time-relaxation-loss}
\end{equation}
where \(f_a\) are chosen observables.
A time-integrated version may also be used when one wants to penalize slow relaxation throughout the whole interval \([0,\tau]\).
For sampling, \(\mathcal R_\tau\) may instead be chosen as a terminal or time-integrated discrepancy between \(\rho_\psi\) and \(\pi\), such as relative entropy, Wasserstein distance, or total variation distance.

We now describe the adjoint form of the critic.
This derivation is not needed for the scalar finite-difference experiment below, but it shows what the continuum actor--critic gradient computes.
For a fixed parameter value \(\psi\), let \(\rho_\psi(\theta,t)\) solve
\begin{equation}
  \partial_t\rho_\psi
  =
  \mathcal{L}_\psi^\dagger\rho_\psi,
  \qquad
  \rho_\psi(\theta,0)=\rho_0(\theta),
  \label{eq:forward-fp-psi}
\end{equation}
where \(\mathcal{L}_\psi\) is the backward generator of the dynamics with gauge drift \(c_\psi\).
We write
\begin{equation}
  \mathcal{L}_\psi f
  =
  \mathcal{L}_0 f
  +
  c_\psi\cdot\nabla f,
  \label{eq:backward-generator-gauge}
\end{equation}
where \(\mathcal{L}_0\) is the baseline generator without the learned gauge.

Consider the terminal observable loss
\begin{equation}
  \Phi(\rho_\psi(\tau))
  =
  \frac{1}{2}
  \sum_a
  w_a
  \left[
    \rho_\psi(\tau)(f_a)-\pi(f_a)
  \right]^2,
  \label{eq:terminal-observable-loss}
\end{equation}
with
\begin{equation}
  \rho_\psi(t)(f_a)
  =
  \int_{\mathbb{R}^d}
  f_a(\theta)\rho_\psi(\theta,t)\,\dd\theta,
  \qquad
  \pi(f_a)
  =
  \int_{\mathbb{R}^d}
  f_a(\theta)\pi(\theta)\,\dd\theta.
\end{equation}
The full finite-time objective is
\begin{equation}
  \mathcal{J}_\lambda[\psi]
  =
  \Phi(\rho_\psi(\tau))
  +
  \int_0^\tau
  \int_{\mathbb{R}^d}
  \rho_\psi(\theta,t)
  r_\psi(\theta)
  \,\dd\theta\,\dd t,
  \qquad
  r_\psi(\theta)
  =
  \frac{\lambda}{2}|c_\psi(\theta)|^2.
  \label{eq:finite-time-objective-with-running-cost}
\end{equation}
The first term measures finite-time relaxation of the chosen observables, while the running cost \(r_\psi\) penalizes the gauge force.

To differentiate \(\mathcal{J}_\lambda\), first vary the forward equation.
A perturbation \(\delta\psi\) induces \(\delta c_\psi\) and \(\delta\rho\), with
\begin{equation}
  \partial_t\delta\rho
  =
  \mathcal{L}_\psi^\dagger\delta\rho
  +
  \delta\mathcal{L}_\psi^\dagger\rho_\psi,
  \qquad
  \delta\rho(\theta,0)=0.
  \label{eq:linearized-fp}
\end{equation}
Since the parameter enters through the drift \(c_\psi\),
\begin{equation}
  \delta\mathcal{L}_\psi f
  =
  \delta c_\psi\cdot\nabla f,
  \qquad
  \delta\mathcal{L}_\psi^\dagger\rho_\psi
  =
  -\partial_i
  \left[
    \rho_\psi\,\delta c_{\psi,i}
  \right].
  \label{eq:variation-generator}
\end{equation}

The adjoint critic \(a(\theta,t)\) is defined backward in time by
\begin{equation}
  -\partial_t a
  =
  \mathcal{L}_\psi a
  +
  r_\psi,
  \qquad
  a(\theta,\tau)
  =
  \frac{\delta\Phi}{\delta\rho_\tau}(\theta).
  \label{eq:finite-time-critic}
\end{equation}
For the terminal observable loss \eqref{eq:terminal-observable-loss}, the terminal condition is
\begin{equation}
  a(\theta,\tau)
  =
  \sum_a
  w_a
  \left[
    \rho_\psi(\tau)(f_a)-\pi(f_a)
  \right]
  f_a(\theta).
  \label{eq:finite-time-critic-terminal}
\end{equation}
The source term \(r_\psi\) appears in \eqref{eq:finite-time-critic} because the objective contains the time-integrated control cost.

Using \eqref{eq:linearized-fp} and integrating by parts in time and space, with vanishing boundary terms, one obtains
\begin{equation}
  \delta\mathcal{J}_\lambda
  =
  \int_0^\tau
  \int_{\mathbb{R}^d}
  \rho_\psi
  \,
  \delta c_\psi
  \cdot
  \nabla a
  \,\dd\theta\,\dd t
  +
  \int_0^\tau
  \int_{\mathbb{R}^d}
  \rho_\psi
  \,
  \delta r_\psi
  \,\dd\theta\,\dd t.
  \label{eq:variation-objective}
\end{equation}
Since
\[
  r_\psi=\frac{\lambda}{2}|c_\psi|^2,
  \qquad
  \delta r_\psi
  =
  \lambda c_\psi\cdot\delta c_\psi,
\]
the gradient with respect to a parameter component \(\psi_k\) is
\begin{equation}
  \frac{\partial\mathcal{J}_\lambda}{\partial\psi_k}
  =
  \int_0^\tau
  \int_{\mathbb{R}^d}
  \rho_\psi(\theta,t)
  \,
  \partial_{\psi_k}c_\psi(\theta)
  \cdot
  \left[
    \nabla a(\theta,t)
    +
    \lambda c_\psi(\theta)
  \right]
  \dd\theta\,\dd t.
  \label{eq:finite-time-gauge-gradient}
\end{equation}

Thus the computation proceeds in three steps.
First, solve or sample the forward dynamics \eqref{eq:forward-fp-psi} to obtain \(\rho_\psi\).
Second, solve the backward adjoint equation \eqref{eq:finite-time-critic}.
Third, use \eqref{eq:finite-time-gauge-gradient} to update the actor parameter:
\begin{equation}
  \psi_{n+1}
  =
  \psi_n
  -
  \alpha_n
  \nabla_\psi
  \mathcal{J}_\lambda[\psi_n].
  \label{eq:actor-update-gauge}
\end{equation}
In this sense, \(A_\psi\) is the actor, because it determines the gauge drift \(c_\psi\), while \(a(\theta,t)\) is the critic, because it evaluates how changes in the drift affect the finite-time objective.

\subsection{Ohzeki--Ichiki Dynamics and the HMC Limit}

We now connect the gauge formulation to the Ohzeki--Ichiki construction.
Let the state variable be
\[
  \theta=(x,y)\in\mathbb{R}^{2d},
\]
where \(x\in\mathbb{R}^d\) is the original variable and \(y\in\mathbb{R}^d\) is an auxiliary variable.
Consider the separable energy
\begin{equation}
  H(x,y)=H_x(x)+H_y(y),
\end{equation}
and the target density
\begin{equation}
  \pi(x,y)
  =
  \frac{1}{Z}
  \exp\left[
    -\frac{H(x,y)}{T}
  \right].
\end{equation}
The generalized Ohzeki--Ichiki dynamics is
\begin{align}
  \dd x_i
  &=
  \left[
    -\partial_{x_i}H_x(x)
    +
    \gamma\,\partial_{y_i}H_y(y)
  \right]\dd t
  +
  \sqrt{2T}\,\dd W^x_i,
  \\
  \dd y_i
  &=
  \left[
    -\partial_{y_i}H_y(y)
    -
    \gamma\,\partial_{x_i}H_x(x)
  \right]\dd t
  +
  \sqrt{2T}\,\dd W^y_i.
  \label{eq:generalized-oi}
\end{align}
Here \(T>0\) is the temperature, \(\gamma\ge 0\) is the strength of the nonreversible force, and \(W^x,W^y\) are independent Brownian motions.

Introduce the constant antisymmetric matrix
\begin{equation}
  S
  =
  \begin{pmatrix}
    0&I\\
    -I&0
  \end{pmatrix},
  \qquad
  S^{\mathsf T}=-S.
\end{equation}
Then \eqref{eq:generalized-oi} can be written compactly as
\begin{equation}
  \dd\theta_t
  =
  \left[
    -\nabla H(\theta_t)
    +
    \gamma S\nabla H(\theta_t)
  \right]\dd t
  +
  \sqrt{2T}\,\dd W_t.
  \label{eq:oi-compact}
\end{equation}
The added drift is
\begin{equation}
  c_\gamma(\theta)
  =
  \gamma S\nabla H(\theta).
\end{equation}
It preserves the target density because
\begin{equation}
  \partial_i\{\pi(\theta)c_{\gamma,i}(\theta)\}
  =
  \gamma\,
  \partial_i
  \left[
    \pi(\theta)S_{ij}\partial_jH(\theta)
  \right]
  =
  0.
\end{equation}
Indeed, the term \(S_{ij}\partial_i\partial_jH\) vanishes by antisymmetry of \(S\), while
\(S_{ij}\partial_iH\,\partial_jH\) also vanishes because it contracts an antisymmetric matrix with a symmetric product.
Thus the Ohzeki--Ichiki force is a constant-antisymmetric-gauge current in the notation of \eqref{eq:antisymmetric-gauge}.
Equivalently, it is generated by the antisymmetric tensor
\begin{equation}
  A_\gamma(\theta)
  =
  -\gamma T S,
\end{equation}
since
\begin{equation}
  \frac{1}{\pi}
  \partial_j
  \left[
    A_{\gamma,ij}\pi
  \right]
  =
  \gamma S_{ij}\partial_jH
\end{equation}
up to the sign convention chosen in \eqref{eq:antisymmetric-gauge}.
This is a symplectic nonreversible flow in arbitrary dimension.
For \(d=1\), the state space is two-dimensional and the current appears as a rotation in the \((x,y)\) plane.

Ichiki and Ohzeki showed that the large-\(\gamma\) limit connects this dynamics to Hamiltonian dynamics \cite{IchikiOhzeki2021}.
After rescaling time by the strength of the nonreversible force, the dominant part of \eqref{eq:oi-compact} becomes
\begin{equation}
  \frac{\dd x_i}{\dd s}
  =
  \partial_{y_i}H_y(y),
  \qquad
  \frac{\dd y_i}{\dd s}
  =
  -\partial_{x_i}H_x(x).
  \label{eq:hmc-limit}
\end{equation}
For
\[
  H_y(y)=\sum_i\frac{y_i^2}{2m_i},
\]
these are Hamilton's equations with masses \(m_i\).
Thus Hamiltonian Monte Carlo arises as the infinite-strength detailed-balance-breaking limit of the Ohzeki--Ichiki dynamics.

This viewpoint also explains why the HMC limit is not automatically optimal.
Hamiltonian flow transports probability rapidly along energy surfaces, but it preserves the energy \(H(x,y)\).
Transitions between energy surfaces must therefore come from momentum resampling, noise, friction, or Metropolis--Hastings correction
\cite{Metropolis1953,Hastings1970,Duane1987,HoffmanGelman2014}.
A finite nonreversible Langevin gauge combines two effects: symplectic motion along energy surfaces and stochastic diffusion across them.
Once finite-time transport performance, control cost, discretization stability, and possible Metropolis rejection are included, the optimal gauge strength need not be \(\gamma=\infty\).

In the terminology of this paper, the Ohzeki--Ichiki family is a one-parameter subfamily of admissible antisymmetric gauges.
The learned-gauge problem asks whether a finite gauge field \(A^\star\), or in this restricted case a finite strength \(\gamma^\star\), gives a better balance than either reversible Langevin dynamics \((\gamma=0)\) or the Hamiltonian limit \((\gamma=\infty)\).
When the resulting nonreversible Langevin proposal is used for exact sampling after time discretization, it can be combined with a Metropolis correction.
We refer broadly to this finite-gauge Langevin strategy as SLGD, emphasizing Langevin dynamics augmented by a symplectic or nonreversible gauge force rather than the infinite-strength HMC limit.

\subsection{Analytical Gaussian Benchmark}
\label{sec:gaussian-validation}

Before turning to a nonconvex metastable landscape, we consider an exactly solvable convex benchmark.
Let \(x\in\mathbb{R}^d\) and consider the quadratic potential
\begin{equation}
  U(x)
  =
  \frac{1}{2}x^{\mathsf T}Kx,
  \qquad
  K=\Sigma^{-1},
  \label{eq:gaussian-quadratic-potential}
\end{equation}
where \(K\) is symmetric positive definite.
The target density is
\begin{equation}
  \pi(x)
  =
  \frac{1}{Z}
  \exp\left[
    -\frac{1}{2T}x^{\mathsf T}Kx
  \right].
  \label{eq:gaussian-target}
\end{equation}
At \(T=1\), the stationary covariance is \(\Sigma\); for general \(T\), it is \(T\Sigma\).
The reversible Langevin dynamics is the Ornstein--Uhlenbeck process
\begin{equation}
  \dd x_t
  =
  -Kx_t\,\dd t
  +
  \sqrt{2T}\,\dd W_t.
  \label{eq:reversible-ou}
\end{equation}
If \(K\) has a small eigenvalue, relaxation along the corresponding flat direction is slow.

We now add a constant antisymmetric gauge.
Let \(S^{\mathsf T}=-S\) and define the stationary-density-preserving drift
\begin{equation}
  c_a(x)
  =
  aSKx,
  \label{eq:ou-gauge-drift}
\end{equation}
where \(a\in\mathbb{R}\) is a scalar gauge strength.
This is generated by the antisymmetric tensor potential \(-aTS\) in the convention of \eqref{eq:antisymmetric-gauge}.
The gauge-modified OU process is
\begin{equation}
  \dd x_t
  =
  -M_a x_t\,\dd t
  +
  \sqrt{2T}\,\dd W_t,
  \qquad
  M_a=(I-aS)K.
  \label{eq:gauge-ou}
\end{equation}
The mean \(\mu_a(t)=\mathbb E[x_t]\) obeys
\begin{equation}
  \dot{\mu}_a(t)
  =
  -M_a\mu_a(t).
  \label{eq:ou-mean-equation}
\end{equation}
Thus the relaxation of the mean is determined exactly by the spectrum of the non-Hermitian drift matrix \(M_a\).

In two dimensions, take
\begin{equation}
  K
  =
  \begin{pmatrix}
    k_1&0\\
    0&k_2
  \end{pmatrix},
  \qquad
  0<k_1<k_2,
  \qquad
  S
  =
  \begin{pmatrix}
    0&1\\
    -1&0
  \end{pmatrix}.
\end{equation}
The eigenvalues of \(M_a\) are
\begin{equation}
  \lambda_\pm(a)
  =
  \frac{k_1+k_2}{2}
  \pm
  \sqrt{
    \frac{(k_1-k_2)^2}{4}
    -
    a^2k_1k_2
  }.
  \label{eq:ou-gauge-eigenvalues}
\end{equation}
For \(a=0\), the spectral gap is \(k_1\), the slow flat direction.
When
\begin{equation}
  |a|
  \ge
  a_{\mathrm{crit}}
  :=
  \frac{|k_1-k_2|}{2\sqrt{k_1k_2}}
  =
  \frac{1}{2}
  \left|
    \frac{\sigma_1}{\sigma_2}
    -
    \frac{\sigma_2}{\sigma_1}
  \right|,
  \label{eq:ou-critical-gauge}
\end{equation}
where \(k_i=\sigma_i^{-2}\), the eigenvalues become a complex conjugate pair and
\begin{equation}
  \mathrm{Re}\,\lambda_\pm(a)
  =
  \frac{k_1+k_2}{2}.
\end{equation}
Thus the gauge converts independent dissipative relaxation into a chiral spiral and removes the asymptotic slow-mode bottleneck.
For \(k_1=0.1\) and \(k_2=1\), the asymptotic mean relaxation rate improves from \(0.1\) to \(0.55\), a factor of \(5.5\).

The critical value \(a_{\mathrm{crit}}\), however, is not the same as the optimum of a finite-time regularized objective.
It is the smallest gauge strength at which the asymptotic spectral rate reaches the plateau.
The actual finite optimum depends on the time horizon, the initial condition, the chosen observable, and the force penalty.
This distinction is useful because it is exactly the distinction made throughout this paper: spectral acceleration alone does not select a unique finite gauge.

For the finite-time objective, let
\[
  Q_a(t)
  =
  \mathbb E[x_tx_t^{\mathsf T}]
\]
be the second moment.
It satisfies the Lyapunov equation
\begin{equation}
  \dot Q_a(t)
  =
  -M_aQ_a(t)-Q_a(t)M_a^{\mathsf T}+2TI.
  \label{eq:ou-lyapunov}
\end{equation}
The exact regularized cost used in the benchmark is
\begin{equation}
  \mathcal C_{\lambda}^{(\tau)}(a)
  =
  \int_0^\tau
  \|\mu_a(t)\|^2\,\dd t
  +
  \lambda
  \int_0^\tau
  \operatorname{Tr}
  \left[
    B_a^{\mathsf T}B_aQ_a(t)
  \right]
  \dd t,
  \qquad
  B_a=aSK.
  \label{eq:ou-finite-time-cost}
\end{equation}
The first term measures relaxation of the mean, and the second term is the expected gauge-force cost
\(\mathbb E[|c_a(x_t)|^2]\).
Equations \eqref{eq:ou-mean-equation} and \eqref{eq:ou-lyapunov} make this benchmark deterministic: no Monte Carlo sampling is needed to compute the critic.

In the numerical benchmark, we set
\[
  k_1=0.1,\qquad
  k_2=1,\qquad
  T=1,\qquad
  \tau=10,\qquad
  \lambda=5.0\times10^{-3},
\]
and initialize the mean along the slow direction,
\[
  \mu_0=(1,0)^{\mathsf T}.
\]
The spectral transition occurs at
\[
  a_{\mathrm{crit}}=1.423.
\]
Direct minimization of \eqref{eq:ou-finite-time-cost} gives
\[
  a_\lambda^\star=2.909,
  \qquad
  \mathcal C_{\lambda}^{(10)}(a_\lambda^\star)=1.401.
\]
For comparison,
\[
  \mathcal C_{\lambda}^{(10)}(0)=4.323,
  \qquad
  \mathcal C_{\lambda}^{(10)}(a_{\mathrm{crit}})=2.089.
\]
The finite optimum is larger than \(a_{\mathrm{crit}}\) because the finite-time mean relaxation still benefits from additional rotation until the force penalty balances the gain.
This is not a contradiction with the spectral analysis: \(a_{\mathrm{crit}}\) marks the onset of the complex-spectrum plateau, whereas \(a_\lambda^\star\) minimizes a specified finite-time cost.

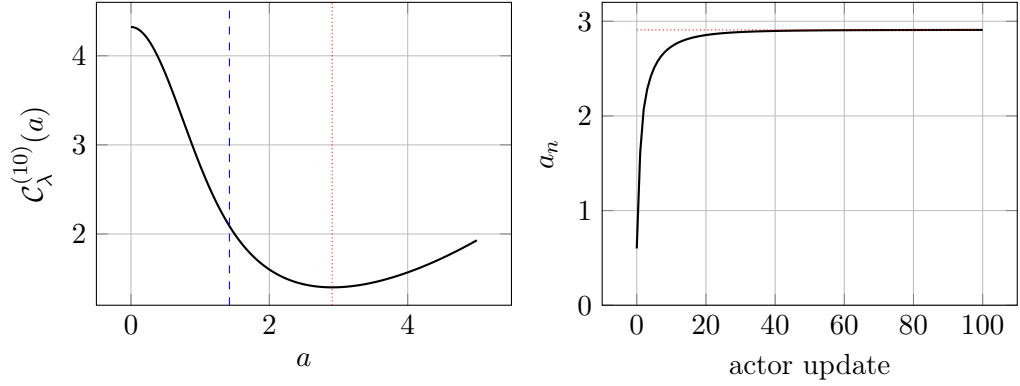
\begin{figure}[tb]
  \centering
  \begin{tikzpicture}
    \begin{groupplot}[
      group style={group size=2 by 1, horizontal sep=1.2cm},
      width=0.43\linewidth,
      height=0.34\linewidth,
      grid=both,
    ]
      \nextgroupplot[
        xlabel={$a$},
        ylabel={$\mathcal C_{\lambda}^{(10)}(a)$},
        ymin=1.2,
        ymax=4.6,
      ]
        \addplot+[thick, mark=none, black] table[x=a,y=objective,col sep=comma] {data/ou_gaussian_cost.csv};
        \addplot+[dashed, mark=none, blue] coordinates {(1.423,1.2) (1.423,4.6)};
        \addplot+[densely dotted, mark=none, red] coordinates {(2.909,1.2) (2.909,4.6)};
      \nextgroupplot[
        xlabel={actor update},
        ylabel={$a_n$},
        ymin=0,
        ymax=3.2,
      ]
        \addplot+[thick, mark=none, black] table[x=iteration,y=a,col sep=comma] {data/ou_gaussian_actor.csv};
        \addplot+[densely dotted, mark=none, red] coordinates {(0,2.909) (100,2.909)};
    \end{groupplot}
  \end{tikzpicture}
  \caption{Finite-gauge learning in the anisotropic Gaussian OU benchmark (color online). In the left panel, the black solid curve is the finite-time objective, the blue dashed vertical line is the spectral threshold \(a_{\mathrm{crit}}\), and the red dotted vertical line is the finite-time optimum \(a_\lambda^\star\simeq2.909\). In the right panel, the black solid curve is the actor trajectory and the red dotted horizontal line is \(a_\lambda^\star\).}
  \label{fig:ou-gaussian-learning}
\end{figure}

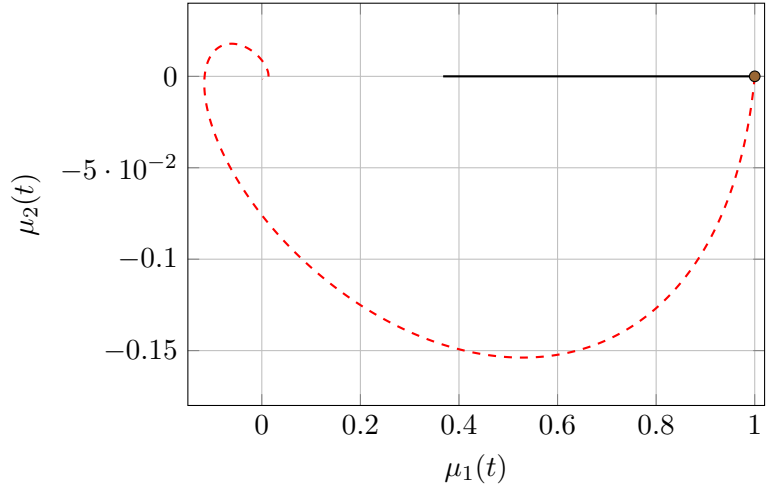
\begin{figure}[tb]
  \centering
  \begin{tikzpicture}
    \begin{axis}[
      width=0.56\linewidth,
      height=0.42\linewidth,
      xlabel={$\mu_1(t)$},
      ylabel={$\mu_2(t)$},
      grid=both,
      xmin=-0.15,
      xmax=1.02,
      ymin=-0.18,
      ymax=0.04,
    ]
      \addplot+[thick, mark=none, black] table[x=mu1_a0,y=mu2_a0,col sep=comma] {data/ou_gaussian_paths.csv};
      \addplot+[thick, mark=none, dashed, red] table[x=mu1_alearned,y=mu2_alearned,col sep=comma] {data/ou_gaussian_paths.csv};
      \addplot+[only marks, mark=*, black] coordinates {(1,0)};
    \end{axis}
  \end{tikzpicture}
  \caption{Mean trajectory in the anisotropic Gaussian OU benchmark from \(\mu_0=(1,0)^{\mathsf T}\) (color online). The black solid curve is the reversible trajectory \(a=0\), the red dashed curve is the learned gauge \(a=2.909\), and the marker denotes the initial mean. The reversible trajectory decays only along the slow direction, while the learned nonreversible gauge bends the flow into a chiral path and accelerates relaxation.}
  \label{fig:ou-gaussian-path}
\end{figure}

We then apply the same scalar actor update used later in the double-well experiment.
Starting from \(a_0=0.6\), using a centered finite-difference critic with \(\epsilon=0.05\) and step size
\[
  \alpha_n=\frac{0.5}{1+0.05n},
\]
100 actor updates give
\[
  a_{\mathrm{learned}}=2.909,
\]
matching the finite-time optimum \(a_\lambda^\star\) computed from the Lyapunov equations.
Figure~\ref{fig:ou-gaussian-learning} shows the objective curve and the actor trajectory.
Figure~\ref{fig:ou-gaussian-path} shows the mean trajectory: the reversible process remains on the slow axis, while the learned gauge produces a chiral relaxation path toward the origin.

\subsection{Double-Well Validation by Forward Simulation}

As a toy validation, we use the double-well potential
\begin{equation}
  U(x)=\frac{x^4}{4}-\frac{x^2}{2}.
\end{equation}
We introduce an auxiliary variable \(y\) and define
\begin{equation}
  H(x,y)=U(x)+\frac{1}{2}y^2,
  \qquad
  S=
  \begin{pmatrix}
    0&1\\
    -1&0
  \end{pmatrix}.
\end{equation}
The one-parameter Ohzeki--Ichiki gauge family is
\begin{equation}
  c_\gamma(x,y)
  =
  \gamma S\nabla H(x,y)
  =
  \gamma
  \begin{pmatrix}
    y\\
    -U'(x)
  \end{pmatrix}.
  \label{eq:oi-gauge-family}
\end{equation}
Thus the learned actor is the scalar gauge strength \(\gamma\), which is a one-dimensional restriction of the general antisymmetric gauge field \(A_\psi\).

For a given \(\gamma\), we simulate the forward Langevin dynamics
\begin{equation}
  \dd
  \begin{pmatrix}
    x_t\\
    y_t
  \end{pmatrix}
  =
  \left[
    -\nabla H(x_t,y_t)
    +
    c_\gamma(x_t,y_t)
  \right]\dd t
  +
  \sqrt{2T}\,\dd W_t.
  \label{eq:double-well-sde}
\end{equation}
The target stationary density is symmetric in \(x\), so \(\mathbb{E}_\pi[x]=0\).
We therefore measure relaxation from the right well by the finite-time observable error \(|\mathbb{E}_\gamma[x_t]|\).

For a time horizon \(\tau\), the rollout objective is
\begin{equation}
  \mathcal{C}_{\lambda}^{(\tau)}(\gamma)
  =
  \int_0^\tau
  |\mathbb{E}_\gamma[x_t]|
  \,\dd t
  +
  \lambda
  \int_0^\tau
  \mathbb{E}_\gamma
  \left[
    |c_\gamma(x_t,y_t)|^2
  \right]
  \dd t.
  \label{eq:numerical-cost}
\end{equation}
Equivalently,
\begin{equation}
  |c_\gamma(x,y)|^2
  =
  \gamma^2
  \left[
    y^2+\{U'(x)\}^2
  \right].
\end{equation}
The first term measures finite-time relaxation, and the second term penalizes the strength of the nonreversible force.

We optimize \(\gamma\) directly from forward simulations.
At actor update \(n\), we evaluate
\[
  \widehat{\mathcal{C}}_{\lambda}^{(\tau)}(\gamma_n+\epsilon),
  \qquad
  \widehat{\mathcal{C}}_{\lambda}^{(\tau)}(\gamma_n-\epsilon)
\]
with common random numbers and form the finite-difference estimate
\begin{equation}
  \widehat{g}_n
  =
  \frac{
    \widehat{\mathcal{C}}_{\lambda}^{(\tau)}(\gamma_n+\epsilon)
    -
    \widehat{\mathcal{C}}_{\lambda}^{(\tau)}(\gamma_n-\epsilon)
  }{2\epsilon}.
\end{equation}
The scalar actor update is
\begin{equation}
  \gamma_{n+1}
  =
  \Pi_{[0,\gamma_{\max}]}
  \left[
    \gamma_n-\alpha_n\widehat{g}_n
  \right].
  \label{eq:scalar-actor-update}
\end{equation}
Thus the numerical procedure is simply: simulate the dynamics forward, estimate the expectations in \eqref{eq:numerical-cost}, compute a finite-difference gradient, and update \(\gamma\).

In the numerical experiment, we set
\[
  T=0.45,\qquad
  \tau=24,\qquad
  \lambda=3.0\times10^{-3},
  \qquad
  \epsilon=0.15,
  \qquad
  \gamma_{\max}=6.
\]
The actor is initialized at \(\gamma_0=0.7\) and updated for 160 iterations with
\[
  \alpha_n=\frac{0.25}{1+0.10n}.
\]
Training rollouts use time step \(2.0\times10^{-3}\) and 800 trajectories.
Validation rollouts use time step \(1.0\times10^{-3}\) and 2500 trajectories with independent random seeds.
All trajectories start from \(x(0)=1\), and \(y(0)\) is sampled from the Gaussian equilibrium marginal \(N(0,T)\).

With these settings, the learned value is
\begin{equation}
  \gamma_{\mathrm{learned}}=2.006.
\end{equation}

For validation, the learned gauge gives
\[
  \int_0^{24}|\mathbb{E}[x_t]|\dd t=1.243,
  \qquad
  \mathcal{C}_{\lambda}^{(24)}=1.589.
\]
This is slightly below the fixed \(\gamma=2\) baseline, which gives \(1.248\) and \(1.592\), respectively.
The stronger fixed \(\gamma=3\) baseline has a smaller raw relaxation area, \(1.016\), but the larger force penalty raises its regularized objective to \(1.787\).
Thus the optimized finite gauge is not the strongest gauge; it is the one that best balances relaxation and control cost under \(\mathcal{C}_\lambda\).

\begin{figure}[tb]
  \centering
  \begin{tikzpicture}
    \begin{axis}[
      width=0.82\linewidth,
      height=0.48\linewidth,
      xlabel={time},
      ylabel={$\mathbb{E}[x_t]$},
      grid=both,
      ymin=-0.08,ymax=1.05,
    ]
      \addplot+[thick, mark=none, black] table[x=t,y=mean_x_gamma0,col sep=comma] {data/relaxation.csv};
      \addplot+[thick, mark=none, blue] table[x=t,y=mean_x_gamma1,col sep=comma] {data/relaxation.csv};
      \addplot+[thick, mark=none, red] table[x=t,y=mean_x_gammalearned,col sep=comma] {data/relaxation.csv};
      \addplot+[thick, mark=none, densely dotted, green!60!black] table[x=t,y=mean_x_gamma3,col sep=comma] {data/relaxation.csv};
      \addplot+[thick, mark=none, dashed, orange!80!black] table[x=t,y=mean_x_hmc,col sep=comma] {data/relaxation.csv};
    \end{axis}
  \end{tikzpicture}
  \caption{Relaxation of the mean position in the double-well model (color online). The black, blue, and red solid curves show \(\gamma=0\), \(\gamma=1\), and the learned finite gauge \(\gamma=2.006\), respectively; the green dotted curve shows \(\gamma=3\), and the orange dashed curve shows HMC. The learned finite gauge improves the regularized objective \(\mathcal{C}_\lambda\), while the stronger fixed gauge \(\gamma=3\) is faster in raw relaxation but has a larger control cost.}
  \label{fig:relaxation}
\end{figure}
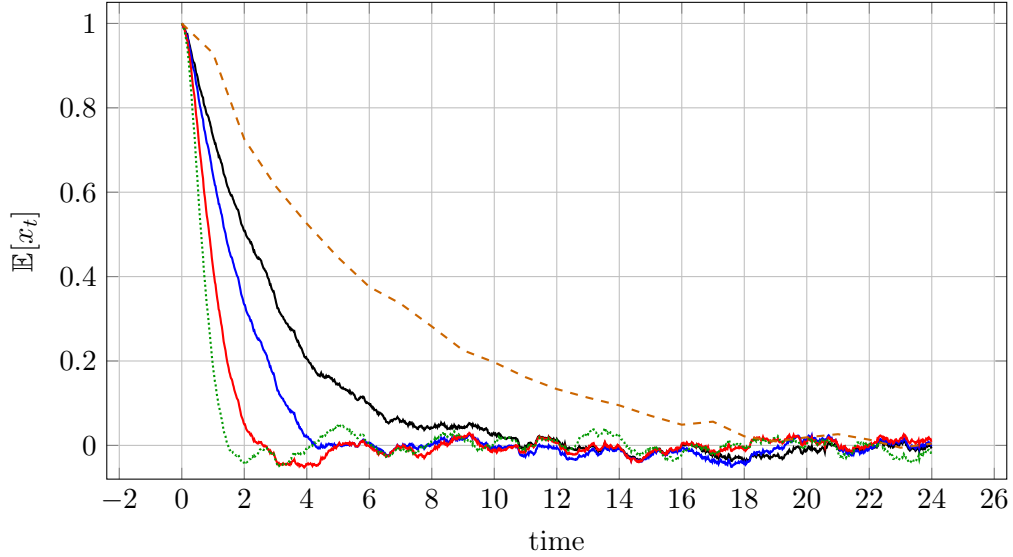

Figure~\ref{fig:relaxation} shows the relaxation of \(\mathbb{E}[x_t]\).
The learned nonreversible gauge accelerates reversible Langevin dynamics and the fixed \(\gamma=1\) baseline.
The fixed \(\gamma=3\) curve relaxes faster in the raw observable, but its control cost is larger.
HMC is included as a reference for the Hamiltonian limit; in this setting its relaxation per unit physical time is slower because transitions between energy shells depend on momentum resampling.
For the HMC reference, we use leapfrog step size \(0.025\), 40 leapfrog steps per proposal, full momentum refresh between proposals, and Metropolis correction.
The corresponding proposal time is \(1.0\), and the observed acceptance rate is \(0.9999\).

\section{Conclusion and Outlook}

We developed an operator-theoretic framework for nonreversible Fokker--Planck dynamics with stationary-density-preserving gauge fields.
In the reversible case, a similarity transformation maps the Fokker--Planck operator to a Witten-Laplacian-type supersymmetric Hamiltonian.
The invariant density becomes the ground state, while relaxation is controlled by the low-lying excited spectrum.
Nonreversible gauges generated by antisymmetric tensor fields preserve this ground state but deform the excited spectrum, producing circulating probability currents and, in general, complex relaxation eigenvalues.

This perspective unifies several structures that are often treated separately.
Mini-batch stochastic gradients appear as effective diffusion tensors or temperatures; adaptive gradient methods appear as metric choices; and detailed-balance-breaking forces appear as gauge currents.
The Ohzeki--Ichiki force is a concrete symplectic gauge in this language, and its infinite-strength limit connects to Hamiltonian dynamics and HMC.
However, the strongest nonreversible force is not automatically optimal.
The continuous-time spectral gap may improve or saturate with increasing gauge strength, while finite-time transport, force cost, discretization stability, and Metropolis rejection impose additional constraints.

We therefore formulated finite-gauge selection as a regularized relaxation problem.
The anisotropic Gaussian OU benchmark gives an exactly solvable test: the spectral transition occurs at \(a_{\mathrm{crit}}=1.423\), whereas the chosen finite-time objective selects \(a_\lambda^\star=2.909\), and the actor update recovers this Lyapunov-equation optimum.
The double-well benchmark shows the same principle in a nonconvex metastable setting: the learned value \(\gamma_{\mathrm{learned}}=2.006\) slightly improves the regularized objective relative to the fixed \(\gamma=2\) baseline, while the stronger \(\gamma=3\) gauge relaxes faster in the raw observable but pays a larger control cost.
Thus finite nonreversible gauges are constrained optima, not simply approximations to the HMC limit.

Several directions remain open.
First, the spectral effect of state-dependent diffusion tensors and adaptive metrics should be characterized more rigorously in high dimension, for example using hypocoercivity or curvature-based methods \cite{BakryEmery1985}.
Second, practical optimizers such as Adam should be analyzed as extended-state Fokker--Planck dynamics, separating metric adaptation from genuine stationary currents.
Third, scalable parameterizations of antisymmetric tensor fields \(A_\psi\) are needed, together with rollout-based or adjoint-based critics for finite-time objectives.
Fourth, finite-gauge Langevin methods with optional Metropolis correction should be benchmarked against SGLD, HMC, nonreversible Langevin samplers, learned MCMC proposals, and Adam-type optimizers \cite{SongZhaoErmon2017,LevyHoffmanSohlDickstein2018}.

The central physical message is that nonreversible gauges provide a controlled way to deform the excited spectrum of a Fokker--Planck operator while preserving its stationary state.
This makes gauge learning a problem of physically constrained spectral design for stochastic dynamics.

\begin{acknowledgments}

The author also acknowledges financial support from the Cross-ministerial Strategic Innovation Promotion Program (SIP) of the Cabinet Office (No. 23836436).

\end{acknowledgments}

\bibliography{references}

\end{document}